\documentclass[preprint]{aastex}

\shorttitle{O VI in High Velocity Clouds}
\shortauthors{Sembach et al.}

\begin{document}

\newcommand{\kms}{\,km\,s$^{-1}$}     
\newcommand{\updag}{$^\dagger$}     

\title{FUSE Observations of \ion{O}{6} in High Velocity Clouds}

\author{K.R.~Sembach\altaffilmark{1}, 
        B.D.~Savage\altaffilmark{2}, 
	J.M.~Shull\altaffilmark{3}, 
 	E.B.~Jenkins\altaffilmark{4}, 
	E.M.~Murphy\altaffilmark{1}, 
	D.G.~York\altaffilmark{5}, 
	T.~Ake\altaffilmark{1},
	W.P.~Blair\altaffilmark{1},
	A.F.~Davidsen\altaffilmark{1}, 
	S.D.~Friedman\altaffilmark{1}, 
	B.K.~Gibson\altaffilmark{3},
	J.W.~Kruk\altaffilmark{1}, 
	H.W.~Moos\altaffilmark{1},
	W.R.~Oegerle\altaffilmark{1},
	D.~Sahnow\altaffilmark{1},
	and G. Sonneborn\altaffilmark{6}}
\altaffiltext{1}{Department of Physics \& Astronomy, The Johns 
	Hopkins University, Baltimore, MD  21218}
\altaffiltext{2}{Department of Astronomy, University of Wisconsin, Madison, 
	WI  53706}
\altaffiltext{3}{CASA, Department of Astrophysical and Planetary Sciences,
University of Colorado, Boulder, CO  80309}
\altaffiltext{4}{Princeton University Observatory, Princeton, NJ 08544}
\altaffiltext{5}{Astronomy \& Astrophysics Center, University of Chicago, 
	Chicago, IL 60637}
\altaffiltext{6}{NASA/GSFC, Greenbelt, MD 20771}

\begin{abstract}
We have used moderate-resolution (FWHM $\approx$ 25 \kms) spectra of 
AGNs and QSOs observed by the {\it Far Ultraviolet Spectroscopic Explorer} to 
make the first definitive measurements of absorption by hot gas in 
high velocity clouds (HVCs) at large distances from the Galactic plane.  
Seven of the 11 sight lines studied 
exhibit high velocity ($|$V$_{LSR}$$| > 100$ \kms) \ion{O}{6}
$\lambda1031.93$ absorption with log\,N(\ion{O}{6}) $\approx$ 13.79--14.62.  
High velocity \ion{O}{6} absorption is detected in the distant gas of
 \ion{H}{1} HVC Complex~C,
the Magellanic Stream, several HVCs believed to be in the Local Group,
and the outer Galaxy.   The fraction of \ion{O}{6}
in HVCs along the seven sight lines containing high velocity \ion{O}{6}
averages $\sim$30\%, with a full range of $\sim$10--70\%.  
 The \ion{O}{6} detections imply that hot 
(T $\sim 3\times10^5$~K), collisionally-ionized gas is an important
constituent of the HVCs since \ion{O}{6} is difficult to produce by 
photoionization unless the path lengths over which the absorption occurs are
very large ($>$100 kpc).  The association of \ion{O}{6} with 
\ion{H}{1} HVCs in many cases suggests that the \ion{O}{6} may be 
produced at interfaces or mixing layers between the \ion{H}{1} clouds and 
hot, low density gas in the Galactic corona or Local Group.  Alternatively, 
the \ion{O}{6} may originate within cooling regions of hot gas clouds as they 
are accreted onto the Galaxy. 

\end{abstract}


\section{Introduction}

Since their discovery nearly 40 years ago, interstellar high velocity  
clouds (HVCs) have remained enigmatic.  Their origins, fundamental physical
properties, and distances are unknown in most cases (see review by 
Wakker \& van~Woerden 1997). 
Although HVCs are often described 
in terms of their \ion{H}{1} 21\,cm emission and their peculiar 
velocities\footnotemark\ with respect to the motions of the general 
interstellar medium,
the traditional view of HVCs as completely neutral entities has given 
way to one that recognizes the importance of an ionized component to many of
the clouds.  H$\alpha$ emission has been detected toward several large 
HVC complexes (Tufte, Reynolds, \& Haffner 1998; Bland-Hawthorn et al. 1998; 
Weiner \& Williams 1996), and recent absorption line observations 
with the {\it Hubble
Space Telescope} have shown that some HVCs are almost fully ionized 
(Sembach et al. 1995, 1999). 
\footnotetext{Usually, clouds moving in excess of 100 \kms\ with respect to 
the Local Standard of Rest fall into the category of high velocity cloud.}

Understanding the ionization of HVCs is a key step toward a more complete
description of the high velocity gas. The large peculiar 
velocities of HVCs make them excellent candidates for studying hot gas 
within the clouds and their 
surrounding environments. A prime diagnostic of  gas at temperatures 
T~$\sim10^5-10^6$~K is the
\ion{O}{6} $\lambda\lambda1031.93, 1037.62$ doublet, which can be observed
in the spectra of distant QSOs and AGNs with the newly commissioned 
the {\it Far Ultraviolet Spectroscopic Explorer} (FUSE). 

In this Letter we report the first detections of \ion{O}{6} absorption in 
several HVCs located in different regions of the sky. The directions 
considered are listed in Table~1, where we summarize sight line 
characteristics and basic measurements for the HVCs. The reader may 
find information about the overall extent and distribution of \ion{O}{6} in 
the Galactic halo in a companion paper (Savage et al. 2000). 
A description of the FUSE mission can be found in Moos et al. (2000).  

\section{Observations and Data Processing}

The FUSE data for this investigation were obtained during the commissioning
stage of the mission in late 1999.  
Each observation was obtained with the source centered in the 
$30\arcsec\times30\arcsec$ aperture of the LiF1 spectrograph channel. 
Exposure times ranged from 13 ksec (Ton\,S210) to 55 ksec (Mrk\,509) (see 
Savage et al. 2000). 
The time-tagged photon lists were processed through the 
standard FUSE calibration pipeline available at the Johns Hopkins University 
as of November 1999.  The lists were screened for valid data with constraints 
imposed for earth limb angle avoidance and passage through the South 
Atlantic Anomaly.  Corrections for detector backgrounds, Doppler shifts 
caused by spacecraft orbital motions, and geometrical distortions were 
applied (see Sahnow et al. 2000).  
No corrections were made for optical astigmatism aberrations or small 
spectral shifts introduced by thermal effects since the data 
were obtained prior to completion of in-orbit focusing 
activities. 
The primary effect of omitting these processing steps is to degrade the 
spectral resolution slightly.

The  processed data have a nominal spectral resolution of 25 \kms\
(FWHM), with a relative wavelength dispersion solution accuracy of 
$\sim 6$ \kms\ ($1\sigma$).  The zero point of the wavelength 
scale for each observation was determined by registering the 
H$_2$ (6--0) P(3) line at 1031.19\,\AA\ to the velocity of the peak \ion{H}{1}
21\,cm emission for the sight line.  In cases where no H$_2$  absorption
is detected, we compared the velocities of  the strong \ion{Si}{2} 
$\lambda1020.70$ and \ion{O}{1} $\lambda1039.23$ lines  to those observed
for lines of the same species at longer wavelengths.


Fully reduced FUSE LiF1A data covering the 1020--1045\,\AA\ spectral region are
shown for PKS\,2155-304 and Mrk\,509 in Figure~1.  Note the presence of 
negative high velocity \ion{O}{6} 1031.93~\AA\ absorption along both 
sight lines.  \ion{O}{6} $\lambda$1037.62 absorption is 
present in both spectra; however, the HVC components are
blended with \ion{C}{2}$^*$ $\lambda$1037.02. 
There is also weak, narrow \ion{C}{2} $\lambda$1036.34
absorption toward both objects near the velocities of the 
\ion{O}{6} HVCs. Other figures showing the 
\ion{O}{6} HVCs considered herein can be found in Savage et al. 
(2000), Oegerle et al. (2000), and Murphy et al. (2000a).

\section{High Velocity O~VI Measurements}

Continuum normalized intensity profiles for the \ion{O}{6} HVC sight lines 
are presented in Figure~2.  Equivalent widths and column 
densities were computed by direct integration of the \ion{O}{6} 
$\lambda1031.93$ intensity and apparent
column density profiles (Sembach \& Savage 1992).
N$_a$(\ion{O}{6})~[cm$^{-2}$] = 
2.748$\times$10$^{12}$ $\int\tau_a$(v)\,dv, where $\tau_a$(v) is 
the measured optical depth of the 1031.93\,\AA\ line 
at velocity v (in \kms) (Savage \& Sembach 1991).
We have assumed an oscillator strength f = 0.133 (Morton 1991).  
N$_a$(\ion{O}{6}) is a reasonable approximation
to N(\ion{O}{6}) since the HVC lines are both weak and broad. The 
integration ranges lie 
beyond the velocities expected for \ion{O}{6} absorption arising in the 
general environment of the thick disk and low halo ($|$z$|$ $<$ 3 kpc) 
(see Savage et al. 2000).  In Table~1 we list the fraction of total 
\ion{O}{6} along each sight line in HVCs, $f_{HVC}$.  On average,  
$f_{HVC} \sim 30\%$.

The H$_2$ (6--0) P(3) 1031.19 \AA\ and R(4) 1032.35 \AA\ lines occur at 
velocities of --214 \kms\ and +123 \kms, respectively, relative to the 
\ion{O}{6} $\lambda$1031.93 line.  We have modeled the impact of these lines 
on the observed \ion{O}{6} profiles by 
examining other H$_2$ (J = 3 or 4) lines in the (3--0) to (8--0) 
vibrational bands covered by the FUSE LiF1A spectra.  
Our estimates of the strengths and shapes of the P(3) 1031.19 \AA\ and 
R(4) 1032.35 \AA\ lines are shown as heavy solid lines in Figure~2.  The 
values of W$_\lambda$ and log~N(\ion{O}{6}) in Table~1 have had the
illustrated H$_2$ contributions removed.

\section{HVC Identifications}
High velocity (V$_{LSR} < -100$ \kms) \ion{O}{6} is seen along 7 of the 
11 extended sight lines toward AGNs and QSOs observed by FUSE. 
The HVCs have negative velocities, with the exception of a weak HVC detected
toward Ton\,S180 (see Table~1). In many cases, the \ion{O}{6} HVCs can be 
related to \ion{H}{1} HVCs located at large distances ($>$3 kpc) from the 
Galactic plane.  

\subsection{The \ion{C}{4} HVCs: Local Group Clouds}
High velocity \ion{O}{6} absorption toward Mrk\,509 and PKS\,2155-304 occurs
at velocities similar to those of some of the \ion{C}{4} HVCs studied by 
Sembach et al. (1999).  The Mrk\,509 \ion{C}{4} HVCs 
($\langle$V$_{LSR}\rangle$ = --287, --228 \kms) are believed to be located 
in the Local Group outside the Milky Way based upon their ionization 
properties and the very low
thermal pressures (p/k $\sim$ 2 cm$^{-3}$ K) inferred if the clouds
are in photoionization
equilibrium.  They display
strong \ion{C}{4} absorption, with little lower ionization absorption and
no detectable \ion{H}{1} 21\,cm emission.  Furthermore, they show no evidence
of H$\alpha$ emission (Sembach, Bland-Hawthorn, \& Savage 2000).
This ionization pattern is 
characteristic of gas clouds irradiated by extragalactic background radiation.

The \ion{O}{6} HVC absorption toward Mrk\,509 is distributed in a broad
component centered on $\langle$V$_{LSR}\rangle \approx -230$ \kms, with FWHM $\approx$ 
120 \kms. It exhibits a steep decline in strength at velocities 
where the \ion{C}{4} is strongest (i.e., in the --287 \kms\ component).  
Most of the \ion{O}{6} appears to be associated
with the lower velocity \ion{C}{4} HVC at --228 \kms.  
The ratio of \ion{C}{4} to \ion{O}{6} is $<$1 in this cloud and the HVCs 
observed toward PKS\,2155-304.  The observed amount of \ion{O}{6} is more than 
an order of magnitude higher than predicted by the standard photoionization 
model\footnotemark\
described by Sembach et al. (1999), which leads us to conclude that there are 
multiple ionization processes in these HVCs.

\footnotetext{Their photoionization model uses an AGN/QSO spectral energy 
distribution with a mean intensity at the Lyman limit 
J$_0$ = 1$\times$10$^{-23}$ erg cm$^{-2}$ s$^{-1}$ Hz$^{-1}$ 
sr$^{-1}$ (Haardt \& Madau 1996).}

\subsection{Complex C}
Mrk\,876 lies behind HVC Complex~C, which is located more than 3.5 kpc from 
the Galactic plane (van~Woerden et al. 1999).  The low metallicity of  
Complex~C, [S/H]\,$\sim$\,$-0.5$ to $-1.0$, indicates that it may be 
material falling onto the Milky Way rather than material ejected from the 
Galactic disk (Wakker et al. 1999; Gibson et al. 2000).
In the direction of Mrk\,876, Complex~C has two distinct \ion{H}{1} components 
centered on V$_{LSR}$ = $-175$ and $-132$ \kms\ (Murphy et al. 2000a).  
The broad \ion{O}{6} absorption spans these neutral 
components and forms a smooth absorption trough (see Figure~2).
  
The presence of \ion{O}{6} at velocities similar to those of the neutral
tracers of Complex~C suggests that Complex~C contains a substantial
amount of ionized gas.  For a gas in collisonal ionization equilibrium
at T = 3$\times$10$^5$~K with Z $\sim$ 0.1--0.3 Z$_\odot$,
N(\ion{O}{6}) = 1.5$\times10^{14}$ cm$^{-2}$ translates into N(H$^+$) 
$\approx(3-9)\times10^{18}$ cm$^{-2}$, or roughly 10--30\% of 
the observed \ion{H}{1} column density of $2.9\times10^{19}$ cm$^{-2}$.

\subsection{Magellanic Stream}
Three of the 7 sight lines exhibiting high velocity \ion{O}{6} lie in 
the general direction of the Magellanic Stream.  Ton\,S210 and Ton\,S180
lie about 10$^\circ$ off the Stream near Complex MSII.  No high  
velocity 21\,cm emission from the Stream is detected toward 
either object; however 21\,cm emission is detected toward Ton\,S210 at
$-196$ \kms\ with a width of 26 \kms\ (Murphy, Sembach, \& Lockman
2000b).  The \ion{H}{1} and \ion{O}{6} HVCs toward Ton\,S210 are
distinct from the nearby Stream material, which has velocities 
V$_{LSR}$ $\gtrsim$ 
--100 \kms\ in this direction.  A sensitive \ion{H}{1} map
of this region indicates that the 21\,cm emission is isolated
(M. Putman, private communication) and resembles
the compact \ion{H}{1} HVCs believed to be located in the Local Group
(Braun \& Burton 1999; but see also Charlton et al. 2000 and Zwaan \& Briggs
2000).

In the direction of NGC\,7469, the high velocity \ion{O}{6} absorption is
related to Magellanic Stream material seen in \ion{H}{1} 21\,cm 
emission near $-350$ \kms\ (Murphy et al. 2000b).  
Ionized Stream gas has been detected previously through its
H$\alpha$ emission (Weiner \& Williams 1996).  Most of that 
emission could be explained by photoionization by starlight (Bland-Hawthorn 
\& Maloney 1999).  However, the presence of 
\ion{O}{6} in the Stream toward NGC\,7469 indicates that hot gas must 
be present since the low gas densities required to produce the observed 
amounts of \ion{O}{6} solely by photoionization requires very
large path lengths ($l \gtrsim 100-200$ kpc $>$ d$_{MS}$).  

\subsection{Outer Galaxy}
Most of the \ion{O}{6} absorption toward H\,1821+643 can be 
attributed to the thick disk and outer Galactic warp (Oegerle 
et al. 2000; Savage et al. 2000).  
The broad, shallow absorption between $-300$ and $-175$ \kms\ 
is probably tracing gas in the most distant portions
of the outer Galaxy.  The limiting velocity of co-rotating interstellar
gas in this direction is roughly --200 \kms. 
Savage, Sembach, \& Lu (1995) noted a very weak \ion{C}{4}
feature with N(\ion{C}{4}) = (1.2$\pm$0.3)$\times10^{13}$ cm$^{-2}$ near
--213 \kms. 
The velocity of the LSR with respect to the velocity centroid of the Local 
Group (l $\approx$ 105$^\circ$, b $\approx$ --8$^\circ$) 
is $\approx -300$ \kms\
(see Mihalas \& Binney 1982), so the projection of this relative
velocity onto the H\,1821+643 sight line is $\approx$ --220 \kms, similar to 
that of the high velocity \ion{O}{6}.  

\section{Discussion}

\ion{O}{6} is difficult to produce by
photoionization (114 eV photons are required).  
The gas density must be so low that path lengths 
exceeding the distance to the Magellanic Stream are necessary
to account for the observed quantities of \ion{O}{6} in the HVCs.
See Sembach et al. (1999) for a discussion of the photoionization models.

Many of the \ion{O}{6} HVCs contain cooler material 
(e.g., Mrk\,876, NGC\,7469), or are in close proximity to \ion{H}{1} HVCs
at similar velocities (e.g., Mrk\,509, PKS\,2155-304, Ton\,S210). 
To produce \ion{O}{6} by shocks requires a shock speed  
of $\approx$\,170 \kms\ (Hartigan, Raymond, \& Hartmann 1987).  The 
observed velocity gradients of the neutral gas in Complex~C (van\,Woerden,
Schwarz, \& Hulsbosch
1985) and the Magellanic Stream (Mathewson \& Ford 1984; Putman \& Gibson
1999) are much smaller than this, so it seems unlikely that cloud-cloud
collisions are responsible for the \ion{O}{6}.  

One possibility for the production of the \ion{O}{6} observed in the HVCs
is that the clouds are moving through a pervasive, hot (T $\sim$ 10$^6$\,K), 
low density (n$_H \lesssim 10^{-4}-10^{-5}$ cm$^{-3}$) Galactic halo or 
Local Group medium.  The existence of such gas has been considered in various 
contexts (see Fabian 1991; Weiner \& Williams 1996; Blitz \& Robishaw 2000; 
Murali 2000). The \ion{O}{6} HVCs have velocities comparable to the adiabatic
sound speed  ($\sim 150$ \kms) for a hot, low density medium.  Thus,
strong shocks are probably not responsible for the \ion{O}{6} production.
Rather, in such a scenario, the \ion{O}{6} would occur in the conductive 
interfaces or turbulently mixed regions of ionized gas between the hot 
medium and the Magellanic Stream or other cooler gas detected in 21\,cm 
emission or ultraviolet absorption.

Alternatively, some of the \ion{O}{6} may be produced within cooling regions 
of hot gas structures associated with the assembly of the Milky Way.  Much
of the baryonic content of the low redshift universe is expected to be 
at temperatures of $10^5-10^7$ K in the vicinity of galaxies and groups of 
galaxies (Cen \& Ostriker 1999).  As the hot gas flows onto galaxies, portions
of it should
cool as the density increases.  Complex~C may represent a relatively
advanced stage of such an accretion, while the HVCs toward 
Mrk\,509 and PKS\,2155-304 would represent an earlier evolutionary stage.
FUSE data for a large number of AGN/QSO sight lines should help to 
distinguish between these various possibilities for the production of the 
\ion{O}{6} HVCs.

\smallskip
This work is based on data obtained for the Guaranteed Time Team by the
NASA-CNES-CSA FUSE mission operated by the Johns Hopkins University.
Financial support has been provided by NASA contract NAS5-32985.

\clearpage
\newpage

\begin{deluxetable}{lccccccccc}
\tablecolumns{10}
\tablewidth{0pt} 
\tablecaption{\ion{O}{6} High Velocity Cloud Sight Lines}
\tablehead{Object & Type & V  & $l$ & $b$ & W$_\lambda$(1031.9) & log N\tablenotemark{a} & $f_{HVC}\tablenotemark{b}$ & Velocity  & HVC\tablenotemark{c} \nl
& & (mag) & ($^\circ$) & ($^\circ$) & (m\AA) &  (cm$^{-2}$) & & (\kms) & ID}
\startdata
Mrk~509 & Syft1 & 13.1 & \phn36.0 & $-$29.9 & 243$\pm$14 & 14.44$\pm^{0.04}_{0.04}$ & 0.35 & $-$380 to \phn$-$95 & 1 \\
PKS~2155-304 & BL~Lac & 13.1 & \phn17.7 & $-$52.3  & 107$\pm$10 & 14.00$\pm^{0.04}_{0.04}$ & 0.33 & $-$300 to \phn$-$80 & 1 \\
Mrk~876 & Syft1 & 15.5 & \phn98.3 & +40.4  & 146$\pm$14 & 14.18$\pm^{0.06}_{0.06}$ & 0.39 & $-$215 to $-$100 & 2 \\
Ton~S180 & Syft1 & 14.3 & 139.0 & $-$85.1 & 156$\pm$21 & 14.29$\pm^{0.07}_{0.10}$ & 0.37 & $-$250 to \phn$-$75 & \nodata \\
         &       &      &       &         & 44$\pm$14  & 13.79$\pm^{0.11}_{0.13}$ & 0.12 & +225 to +310 & \nodata \\
Ton~S210 & QSO & 15.2 & 225.0 & $-$83.2 & 91$\pm$24 & 14.11$\pm^{0.10}_{0.14}$ & 0.23 & $-$280 to $-$130 & 3 \\
NGC~7469 & Syft1 & 13.0 & \phn83.1 & $-$45.5 & 289$\pm$23 & 14.62$\pm^{0.07}_{0.08}$ & 0.74 & $-$400 to $-$100 & 4 \\
H1821+643 & QSO & 14.2 & \phn94.0 & $-$27.4 & 87$\pm$20 & 14.00$\pm^{0.06}_{0.07}$ & 0.21 & $-$330 to $-$175 & 5 \\
\enddata
\tablenotetext{a}{Column density derived by integrating the \ion{O}{6}
1031.926 \AA\ line over the velocity range listed.  The value is corrected for the  (6--0) H$_2$ P(3) 1031.19\AA\
absorption shown in Figure~2.  An additional uncertainty of $\sim\pm$0.04 dex (not listed) 
is appropriate for those observations where H$_2$ absorption has been removed.}
 
\tablenotetext{b}{Ratio of N(\ion{O}{6}) in the high velocity gas within the velocity
range listed compared to total N(\ion{O}{6}) along the sight line.  See Savage et al. (2000)
for values of N(\ion{O}{6}) outside the listed velocity range.}
 
\tablenotetext{c}{HVC Association:  1 = \ion{C}{4}-HVCs detected by Sembach et al. (1999);
2 = Complex C;
3 = Compact \ion{H}{1}-HVC, possibly in the Local Group;
4 = Magellanic Stream;
5 = Distant outer Galaxy.}
 
\end{deluxetable}

\clearpage
\newpage
\noindent
\begin{figure}
\includegraphics{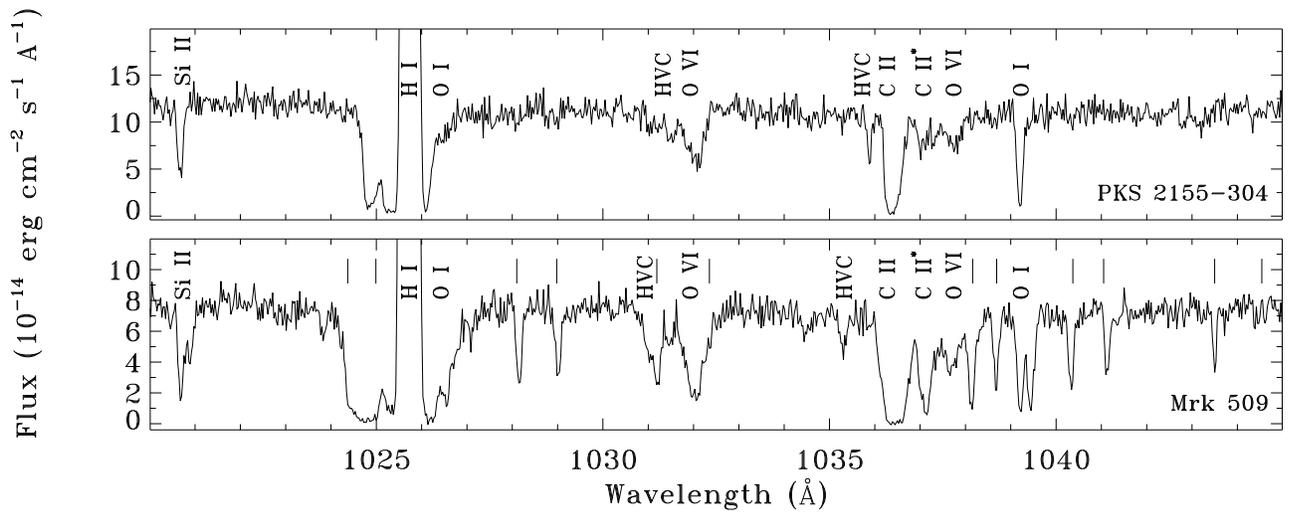}
\caption{FUSE LiF1A data for the 1020--1045\AA\ spectral regions of 
PKS\,2155-304 and Mrk\,509. Ticks 
above the Mrk\,509 spectrum indicate the locations of prominent H$_2$ lines 
in the (5--0) and (6--0) vibrational bands.  Note the presence of \ion{O}{6}
HVCs toward both objects.}
\end{figure}

\clearpage
\newpage
\noindent
\begin{figure}
\includegraphics{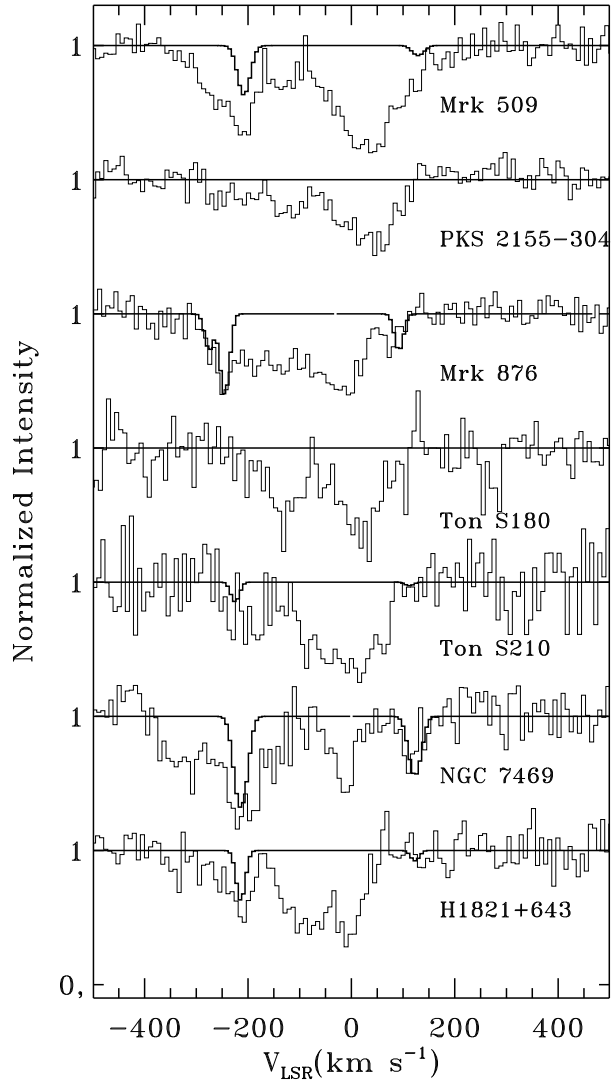}
\caption{Normalized intensity versus LSR velocity for the \ion{O}{6} $\lambda1031.93$ line along the 7 sight lines containing \ion{O}{6} HVCs. The heavy
solid line overplotted on each spectrum is a model of the 
H$_2$ absorption in the (6--0) P(3) $\lambda1031.19$ and R(4) $\lambda1032.35$
lines.}
\end{figure}

\end{document}